\documentclass[]{aa} 
\usepackage{amsmath} 
\usepackage{txfonts}
\usepackage{graphicx} 
\usepackage{url}
\usepackage{natbib} 
\bibpunct{(}{)}{;}{a}{}{,}

\newcommand{\ie}{{\it i.e.}~}

\newcommand{\Ms}{\ensuremath{M_\odot}}

\newcommand{\el}[2]{$\rm{}^{#2}\kern-0.6pt#1$}

\newcommand{\rht}{\ensuremath{r_\text{h}/r_\text{t}}}
\newcommand{\tgcr}{\ensuremath{\tau_{GE}/t_\text{cr}}}
\newcommand{\eps}[1][]{\ensuremath{\epsilon_{#1}}}

\begin{document}

\title{Evolution of two stellar populations in globular clusters}
\subtitle{II. Effects of primordial gas expulsion}


\author{T. Decressin\inst{1} \and H. Baumgardt\inst{1,2} \and
  C. Charbonnel\inst{3,4} \and P. Kroupa\inst{1}}

\offprints{T. Decressin, email: decressin@astro.uni-bonn.de}

\institute{Argelander Institute for Astronomy (AIfA), Auf dem H\"ugel
  71, D-53121 Bonn, Germany
  \and
  Present address: Department of Physics, University of Queensland, Brisbane,
QLD 4072, Australia
  \and
  Geneva Observatory, University of Geneva, 51 ch. des Maillettes, 1290
  Versoix, Switzerland
  \and
  Laboratoire d'Astrophysique de Toulouse-Tarbes, CNRS UMR 5572, Universit\'e de Toulouse, 14 Av. E. Belin, 31400 Toulouse, France
}

\date{Received / Accepted}

\authorrunning{} \titlerunning{}

\abstract{} %
{We investigate the early evolution of two distinct populations of low-mass
  stars in globular clusters under the influence of primordial gas
  expulsion driven by
  supernovae to study if this process can increase the fraction of second
  generation stars at the level required by observations.} %
{We analyse N-body models that take into account the effect of primordial gas
  expulsion. We divide the stars into two populations which mimic the
  chemical and dynamical properties of stars in globular clusters so that
  second generation stars start with a more centrally concentrated distribution.} %
{The main effect of gas expulsion is to eject preferentially first
  generation stars while
  second generation stars remain bound to the cluster. In the most favourable cases
  second generation stars can account for 60\% of the bound stars we see today.
 We also find that at the end of the gas expulsion phase,
  the radial distribution of the two populations is still different,
  so that long-term evolution will further increase the fraction of second
  generation stars.} %
{The large fraction of chemically anomalous stars is readily
    explainable as a second generation of stars formed out of the slow
    winds of rapidly rotating massive stars if globular clusters suffer
    explosive 
    residual gas expulsion for a star formation efficiency of about 0.33.}


\maketitle

\section{Introduction}

Globular clusters are self-gravitating aggregates of tens of thousands to
millions of stars which have survived over a Hubble time. Many observations
show that these objects are composed of (at least) two distinct stellar
populations. The first evidence rests on the chemical analysis that reveals
large star-to-star abundance variations in light elements in all individual
clusters studied so far, while the iron abundance stays constant \citep[for
a review see][]{GrattonSneden2004}. This includes the well-documented
anticorrelations between C-N, O-Na, Mg-Al, Li-Na and F-Na
\citep{Kraft1994,CarrettaBragaglia2006,CarrettaBragaglia2007,CarrettaBragaglia2009,GrattonLucatello2007,PasquiniBonifacio2006,BonifacioPasquini2007,LindPrimas2009}.
This global chemical pattern requires H-burning at high temperatures
around $75\times 10^6$~K
\citep{ArnouldGoriely1999,PrantzosCharbonnel2007}. As the observed chemical
pattern is present in low-mass stars both on the red giant branch (RGB) and
at the turn-off which cannot have reached such high temperatures, the
abundance anomalies must have been inherited at the time of formation of
these stars.

Further indications for multiple populations in individuals GCs comes from
deep photometric studies that have revealed multiple giant branches or
main sequences. In $\omega$~Cen a blue main sequence has been discovered
\citep{BedinPiotto2004} which is presumably related to a high content in He
\citep{PiottoVillanova2005,VillanovaPiotto2007}. A triple main sequence has
been discovered in NGC~2808 \citep{PiottoBedin2007}. A broadening of
  the main sequence of NGC~6752 has also been discovered which can not be
  explained by binary stars \citep{MilonePiotto2010}.
The additional blue
sequences are explainable only by a higher He content of the corresponding
stars which increases the opacity and shifts the effective temperature
towards higher temperatures. He-rich stars are also the progenitors 
  of blue horizontal branch stars seen in many globular clusters \citep[see][]{CaloiD'Antona2005,CaloiD'Antona2006}.
Whereas no direct observational link between abundance anomalies and
He-rich sequences has been found, theoretically this link is easily
understood as abundance anomalies are the main result of H burning to He.
 
These observed properties lead to the conclusion that globular clusters
born from giant gas clouds first form a generation of stars with the same
abundance pattern as field stars.  Then a polluting source enriches the
intracluster-medium with H-burning products out of which a
chemically-different second stellar generation forms with a spread of
  chemical peculiarities. This scheme can
explain at the same time the abundance anomalies in light elements and
He-enrichment.

Two main candidates that reach the right temperature for H-burning have
been proposed to be at the origin of the abundance anomalies
\citep{PrantzosCharbonnel2006}: (a) intermediate mass stars evolving through the
thermal pulses along the asymptotic giant branch (hereafter TP-AGB), and
(b) main sequence massive stars. After being first proposed by
\citet{CottrellDaCosta1981} the AGB scenario has been extensively studied
\citep{VenturaDAntona2001,VenturaDAntona2002,VenturaDAntona2005,VenturaDAntona2005a,VenturaDAntona2005b,VenturaDantona2008a,VenturaDantona2008b,VenturaDantona2009,DenissenkovHerwig2003,KarakasLattanzio2003,Herwig2004a,Herwig2004b,FennerCampbell2004,BekkiCampbell2007,DecressinCharbonnel2009}. In
massive TP-AGB stars ($M\ge4$~\Ms{}), the abundance anomalies are supposed to be
created at the bottom of the convective envelope through hot bottom burning.

On the other hand, as has been suggested by
\citet{WallersteinLeep1987} and \citet{BrownWallerstein1993},
 massive stars can also pollute the
inter-stellar medium (ISM) of a forming cluster
\citep[see][]{Smith2006,PrantzosCharbonnel2006}. In particular
\citet{DecressinMeynet2007} show that fast rotating massive stars (with a
mass higher than $\sim25$~\Ms{}) are good candidates for the
self-enrichment of globular clusters. In the wind of fast rotating massive
stars (WFRMS) scenario, rotationally-induced mixing transports H-burning
products (and hence matter with correct abundance signatures) from the
convective core to the stellar surface, and, provided initial rotation is
high enough, the stars reach break-up velocity while on the main sequence. 
As a result, a mechanical wind is launched from the equator that generates a
disk around the star similar to that of Be stars
\citep[e.g.][]{TownsendOwocki2004}. Later, when He-burning products are
brought to the surface, the star has already lost a high fraction of its
initial mass and angular momentum, so that it no longer rotates at the
break-up velocity. Matter is then ejected through a classical fast
isotropic radiative wind. From the matter lost through the disk, a second
generation of stars may be created with chemical pattern in agreement with
observations. 

In the present paper we mainly focus on the dynamical constraints related
to the WFRMS scenario. However both scenarios are facing a
  similar problem. Within standard assumptions (canonical IMF and
  conservation of the stars in the cluster), the amount of matter lost by
the polluter stars is much smaller than the mass locked into the first
generation low-mass stars. Indeed, if we assume that first generation
massive stars follows the canonical IMF\footnote{The canonical IMF is a two
  part power-law function, $f(m)\propto m^{-\alpha_i}$, with $\alpha_1=1.3$
  for stellar masses $0.08\le m/\Ms \le 0.5$ and $\alpha_2=2.3$ (Salpeter
  value) for $m>0.5$~\Ms \citep{Kroupa2001}.}, stars in the mass range
25--120~\Ms{} account for only 10\% of the mass of the whole first stellar
generation \citep[see][]{DecressinCharbnnel2007} and their slow winds
account for 2.5\% of the mass only. If pollution is due to AGB stars, a
  similar constraint arises: the wind released by stars between 5 and
  6.5~\Ms{}, for which nucleosynthesis agrees with the observations
  according to \citet{VenturaDantona2008a}, represents less than 3\%.
After taking into account the possible dilution of these slow winds with
the pristine gas present in the ISM to explain the observed Li abundance
variation \citep{DecressinCharbnnel2007}, we find that the mass available
to form the second generation low-mass stars compared to the first
generation of low-mass stars is only about 10\%. This is in sharp contrast
with observations, which show that more than half and up to 85\% of the
stars in GCs are second generation stars
\citep{PrantzosCharbonnel2006,CarrettaBragaglia2008}, i.e., which show
  anticorrelations in light elements. Thus a rather extreme reduction of
  the first generation stars relative to the second generation stars is
  needed to reproduce the observations. However massive binaries have
recently been proposed as polluters of the proto-GC by
\citet{deMinkPols2009}. In this case the mass-budget is more favourable as
more slow winds are ejected and more second generation stars are
formed. This could help to reduce the fraction between first and
  second generation stars found in the present paper therefore
supporting a high-mass star pollution scenario.

One
possible way to reconcile the pollution scenario with the observations is
to consider a top-heavy initial mass function (IMF) of first generation
stars.
In the case of pollution by fast-rotating massive stars, an IMF slope
as
flat as 1.55 (compared to the canonical value of 2.3)
is required to reproduce the high number of stars with abundance
anomalies in the cluster NGC~6752 \citep{DecressinCharbnnel2007}, whereas
the AGB scenario requires an even flatter IMF slope
\citep[see][]{PrantzosCharbonnel2006}.  

A second way to reconcile the pollution scenario with observations is to
consider that first generation stars are preferentially lost from the
cluster during its evolution so that an initially relatively small
population of second generation stars can become the dominant population
after several Gyr. To allow this preferential loss of first generation
stars requires the GCs to be initially mass-segregated (i.e., that
more massive stars occupy the central part of the clusters). In this case
the matter released in the disks of massive stars is more concentrated in
the cluster centre, and second generation stars are born in the centre
while first generation stars are present throughout the cluster.

The viability of a self-enrichment scenario by fast-rotating massive stars
has been recently explored by \citet[Paper
I]{DecressinBaumgardt2008}.\defcitealias{DecressinBaumgardt2008}{Paper I}
They have shown that first generation low-mass stars are preferentially
lost from the cluster, which is assumed to be initially in dynamical
equilibrium and mass-segregated, before two-body relaxation induces a
spread of second generation stars and a full mixing of the cluster.
\citet{DErcoleVesperini2008} find similar results with the AGB scenario.
Afterwards, the evolution is smoother and the variation of the fraction of
second generation stars takes longer. Any radial difference between first
and second generation stars is erased after 10-12~Gyr of evolution as the
cluster relaxation time (a few Gyr) is much shorter than the age of the
  clusters.\footnote{The only exception is the GC $\omega$~Cen, for which
  the relaxation time at the center is comparable to its age. Indeed in
  this cluster stars on the blue main sequence (i.e., He-rich) are found
  more centrally concentrated than red main sequence stars
  \citep{VillanovaPiotto2007,BelliniPiotto2009}.} In
  \citetalias{DecressinBaumgardt2008} we show that even if the
  relaxation-driven evaporation increases the fraction of second generation
  (which harbour abundance anomalies) to about 25\%, this ratio remains too
  low to fully explain the observations (between 50--85\%,
  \citealp{CarrettaBragaglia2009}).  The increase of the fraction of second
  generation stars mainly occurs in the early times and points towards the
high sensitivity of the fraction of second generation stars on
cluster dynamics.

In this paper, we aim to quantify the increase of the fraction of
  second generation stars to the total number of low-mass stars by
another dynamical
mechanism not taken into account in the above studies, namely the effect of
primordial gas expulsion (i.e., the fast ejection of the remaining gas
left by star formation after the onset of supernovae). 
Gas expulsion can strongly modify the total binding energy of the 
cluster and can lead to an efficient loss of first generation stars from 
the cluster. We emphasise that we discuss generic properties of gas
  expulsion models based on the simplified assumption that a cluster
  contains only two stellar generations with the same [Fe/H]. Multiple
  populations with different [Fe/H] would require other physical
  mechanisms, whereby notably gas accretion form the surrounding inter
  stellar medium \citep{PflammAltenburgKroupa2009} may play a role, and
  recycling of SN ejecta also \citep{TenorioTagleWunsch2007}.
 In \S~2 we present the N-body models used in this study. Then our
results are discussed in \S~3. In \S~4 we present a complete scenario for the
evolution of GCs and our conclusions are in \S~5. 

\section{Description of analysis}

The results presented in this paper are based on the grid of N-body models
computed by \citet{BaumgardtKroupa2007}, where the effects of primordial gas
expulsion on the dynamics of star clusters were studied. The N-body models are
computed with the \textsc{NBODY4} code \citep{Aarseth1999} and follow the
evolution of $20\,000$ single-mass
 stars. The gas is treated as a spherical
additional potential that is removed gradually in order to change the
total binding energy of the cluster. The initial cluster follows a Plummer
distribution. The cluster evolution is computed for 100 to 150 initial
crossing times so that the cluster can settle into a new equilibrium
configuration and 
two-body relaxation (which acts on a much longer
timescale) is not an important parameter in these models.

\citet{BaumgardtKroupa2007} studied in particular the influence of three
physical parameters on the early cluster dynamics. The first is the star
formation efficiency, \eps{}, given by the ratio between the stellar mass
and the initial mass of the parent gas cloud. This parameter defines the
fraction of gas converted into stars due to star formation. The second
parameter is the ratio between the half-mass radius and the tidal radius,
\rht{}, which quantifies the initial concentration of the cluster and the
strength of the tidal field of the host Galaxy. Finally the third
parameter is the ratio of the timescale for gas expulsion relative to the
crossing time, \tgcr{}. This quantity determines the ability of stars to
adjust their orbital parameters when the potential changes during gas
expulsion. The full grid of models includes variation of the star formation
efficiency between 0.05 and 0.75, of the ratio of the half-mass radius to the
tidal radius between 0.01 and 0.2, and of the ratio of the timescale for gas
expulsion relative to the crossing time between 0 and 10.

As the models of \citet{BaumgardtKroupa2007} take into account only one
stellar population, we use the same method as in
\citetalias{DecressinBaumgardt2008} to split the stars into two populations
according to their specific energy. The stars with the highest binding
energy mimic second generation stars that are more centrally concentrated,
while the other stars are assumed to be members of the first generation. We
choose an initial fraction of second generation stars of 10\% to be
consistent with the pollution by fast rotating massive stars in the case of
a canonical IMF slope. As \citet{BaumgardtKroupa2007} used only
single-mass stars we cannot study in detail the mass dependence of our
results. However the gas expulsion process we investigate here (duration
shorter than 10 crossing times) acts on a much shorter timescale than
two-body relaxation (a few Gyr) which could lead to the preferentially loss
of low-mass stars, and also acts on a shorter timescale than the
lifetime of low-mass stars. So we do not expect our results to
depend much on the mass of the stars. Similarly, the short duration of
  the gas expulsion phase compared to the two-body relaxation timescale 
  allows us to infer results suitable for the study of the early dynamics of
  globular clusters even with the limited number of stars ($20\,000$) in
  the N-body model library.

\section{Results}

\subsection{Analysis of individual models}

\begin{figure*}
  \includegraphics[width=\textwidth]{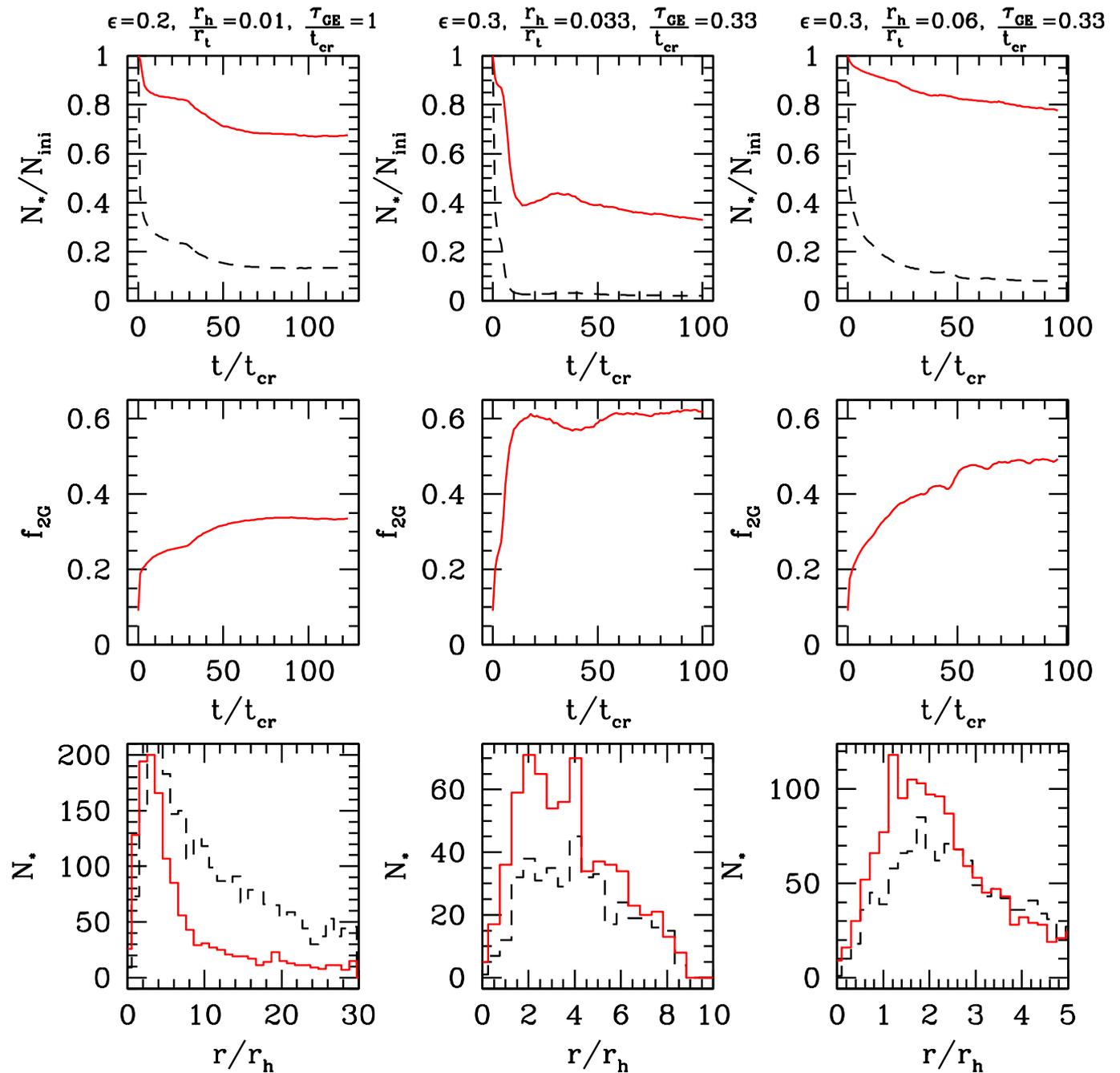}
  \caption{\textit{Top panels:} number fraction of first (dashed lines) and
    second generation (full lines) stars relative to their initial number
    as a function of time. Each line is normalised to its initial number.
    \textit{Middle panel:} fraction of second generation stars bound to the
    cluster as a function of time.  \textit{Bottom panel:} final (at
      100 initial crossing times) radial
    distribution from the cluster centre for the stars of the first (dashed
    lines) and the second (full lines) generation.  Right, central and left
    panels refer to three cases with different initial parameters indicated
    at the top.}
\label{fig:lossGE}
\end{figure*}

\citet{BaumgardtKroupa2007} find that gas expulsion can lead to all
  situations between a cluster totally disrupted and an unaffected cluster,
depending on the parameter values for the gas-expulsion timescale,
  $\tau_\text{GE}$, the star formation efficiency, $\epsilon$, and cluster
  concentration $c$.
In this paper we will concentrate on intermediate cases that
predict a large stellar mass-loss but with a remnant dense core.
In Fig.~\ref{fig:lossGE} we present three such interesting cases.

\textbf{Case~1.}
The left panels show predictions for the gas expulsion parameters $\eps = 0.2$, $c=\rht = 0.01$,
and $\tgcr = 1$, which represent an initially concentrated cluster with a
low star formation efficiency and a long timescale for gas expulsion.  In
the upper left panel we show the evolution of the number of stars of the first
and second
population still bound to the cluster. We use the following criteria
to define if, at a given time, a star is bound to the cluster:
the star needs to be within the tidal radius of the cluster and to have a
negative total energy (sum of the kinetic and potential
energy). Initially some stars have a positive total energy and they move
away from the cluster. However during about the first 10 crossing times the
radius of the stars bound to the cluster expands so that these stars stay
within the cluster tidal radius. During the first 30 crossing times, about 10\%
of the stars are in this situation (positive total energy and still within
the tidal radius of the
cluster) and it depends on the criteria used if they are to be considered as
bound (\citealp{BaumgardtKroupa2007} with only a radius criterium) or unbound (this
study with both a radius and a energy criterium). However after around 30
crossing times, both criteria give the same results so that at the end of the
computation no difference exists between both criteria. The two-phase
decrease of the number of stars comes from the change of an energy
dominating criterion in the early times to a radial criterion that
dominates after 30 crossing times. Using only the radial criterion would
have lead to a smoother decrease of the number of bound stars
\citep[see][]{BaumgardtKroupa2007}.

As expected from the initial radial distribution, case~1 depicts a cluster that loses more of its first generation stars (about 80\%) than second generation
ones (about 12\%) so that the fraction of second generation stars is around
30\% at the end of the computation (middle left panel in
Fig.~\ref{fig:lossGE}). The loss of stars is very pronounced during the first
few crossing times when the lowering of the cluster binding energy is driven by
the gas expulsion.

An interesting point is that the radial distributions (see left bottom
panel in Fig.~\ref{fig:lossGE}) differ at the end of the simulation (about
100 initial crossing times), second generation stars being still more
concentrated than the first generation ones. Thus we can expect that the
fraction of second generation stars will increase further due to the
relaxation driven long-term evaporation of the clusters, as seen in
\citetalias{DecressinBaumgardt2008}. Moreover, the final radius of the
cluster is much larger than initially. Indeed the half-mass radius of first
and second generation stars is 1.0 and 0.6~pc respectively at $t=0$, and
becomes 8.5 and 3.8~pc at the end of the computation.  This strong radial
expansion by a factor 8.4 (first generation) and 6.3 (second generation) is
mainly due to the long timescale for gas expulsion, $\tau_\text{GE}$, that
is similar to the crossing time, allowing stars to adopt new orbital
parameters with wider orbits without being lost from the cluster. It
should be noted that the tidal field is also weak in this case as shown by
the radial extension of the cluster up to about 30 times its initial
half-mass radius.

\begin{figure*}[t]
\includegraphics[width=0.48\textwidth]{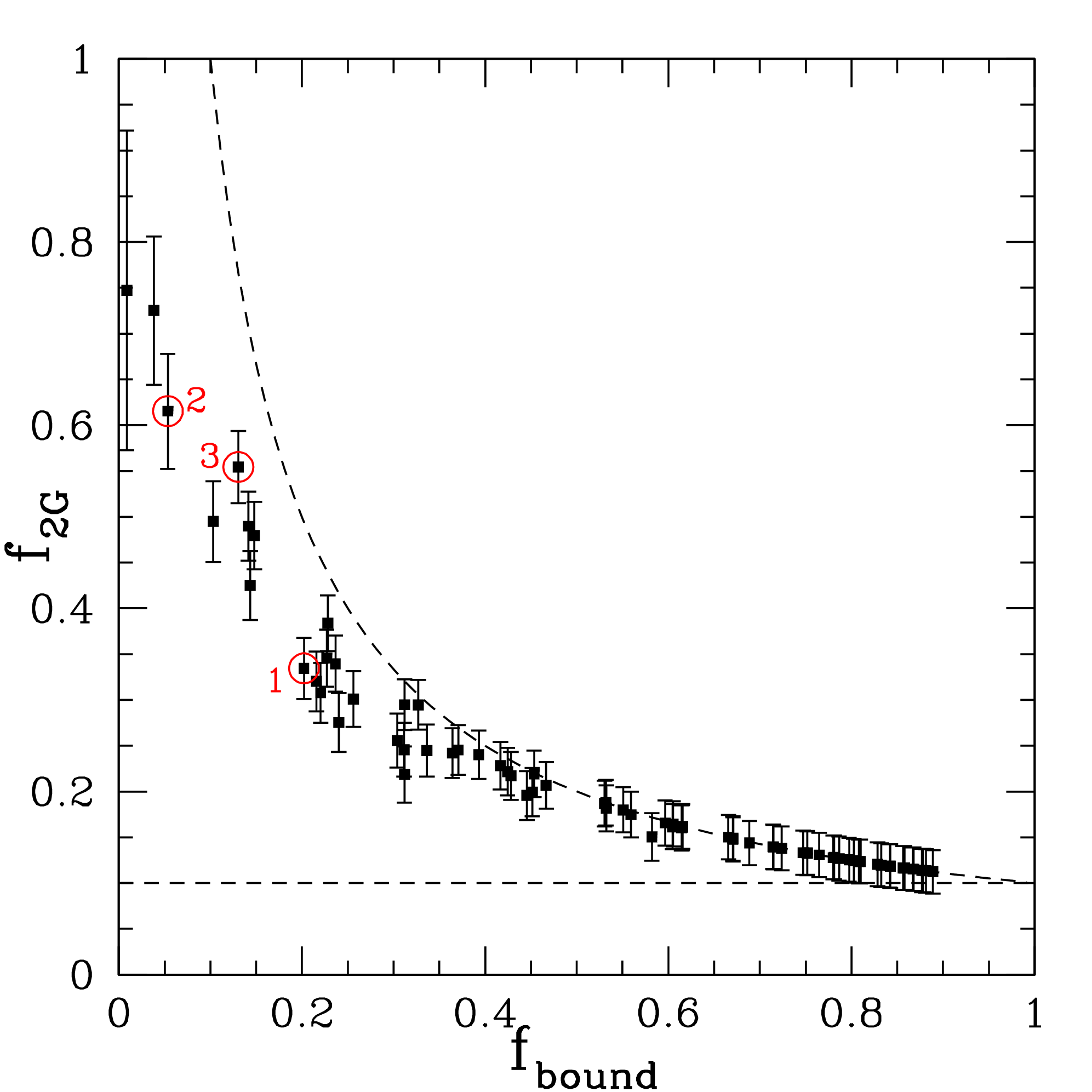}
\includegraphics[width=0.48\textwidth]{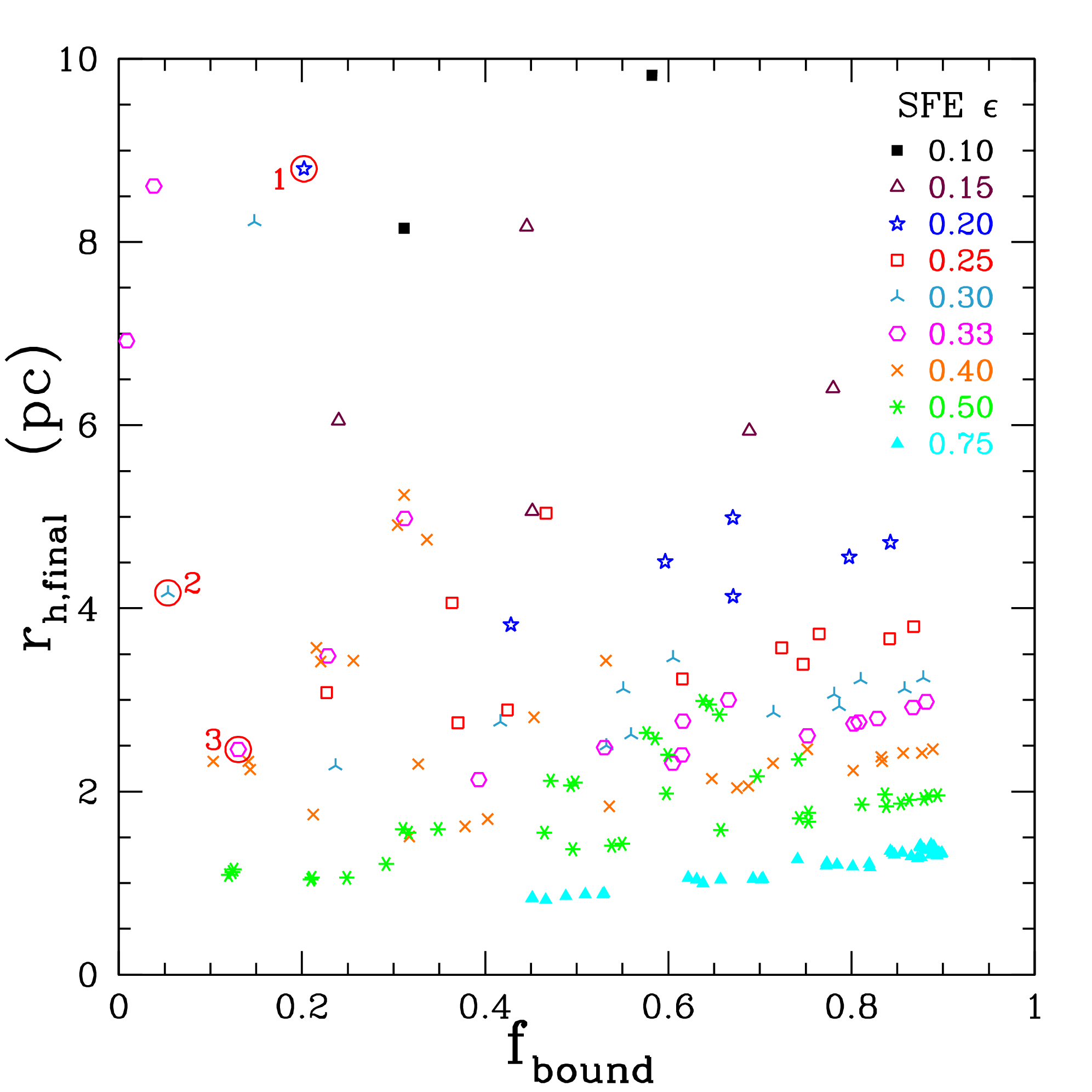}
\caption{\textit{Left:} Fraction of second generation stars as a function of the final
  fraction of bound stars at the end of the computations of
  \citet{BaumgardtKroupa2007}, i.e., after about 100~initial crossing
  times. Dashed lines indicate limiting cases where no second generation
  stars are lost (upper) and no preferential loss of first generation stars
  occurs (lower). Estimates of the statistical errors are also
  included based on the number of first, $N_1$, and second, $N_2$,
  generation stars bound to the cluster. \textit{Right:} Final half-mass
  radius of the cluster as a function of the fraction of bound stars at the
  end of the computation. Clusters with different values of the star formation efficiency, $\epsilon$,
  are indicated with various symbols and colours.
  In both figures, numbered circles indicate the three models presented in
  Fig.~\ref{fig:lossGE} and discussed in \S~3.1.}
\label{fig:all2g}
\end{figure*}

\textbf{Case~2.} The central panels of Fig.~\ref{fig:lossGE} present
the case of a
cluster near total disruption which loses $\sim 95$\% of its first
generation stars due to the gas expulsion process. The initial parameters
are $\eps = 0.30$, $\rht = 0.033$ and $\tgcr = 0.33$. Compared to the
previous case, this model has a smaller gas fraction after star formation
and gas expulsion occurs on a much shorter timescale.  As this cluster
shows a smaller increase of its radius during its evolution the two-phase
decrease of the number of stars is limited to only the first 5
crossing-times. 
The number of bound stars goes through a minimum around 10~crossing times
before increasing during the following 20 crossing-times. This behaviour is
related to the strong ellipsoidal shape that the cluster displays during
its expansion phase, leading to a significant number of stars lying in the
outer part of the major axis of the cluster distribution where they
are outside the tidal radius (and are hence counted as unbound stars). When
the cluster contracts and becomes more spherical, part of these stars
decrease their orbital radius below $r_\text{t}$ and become bound to the
cluster again. At the end of the evolution, the cluster radius has only
increased by a factor 3-4. The central part of the cluster is dominated by
second generation stars (bottom panel) and the fraction of second
generation stars at the end is above 60\%. However due to the low
number of bound stars in the simulation, statistics becomes too poor
to precisely infer cluster properties.

\textbf{Case~3.}  Finally the right panels of Fig.~\ref{fig:lossGE}
correspond to a
model with initial parameters of $\eps = 0.33$, $\rht = 0.06$ and $\tgcr =
0.33$ that also undergoes a strong loss of stars leading to a cluster with
second generation stars counting for half of all cluster members. This case
is particularly interesting as it has a small final half-mass radius which
is only about twice the initial half-mass radius. Second generation stars
dominate at the centre, thus we can expect that the further evolution of
this cluster will
increase the fraction of second generation stars to match the observed
fraction (50--85\%) of stars with anticorrelations in light elements
when taking into account its whole evolution.

\subsection{The whole set of models}

Figure~\ref{fig:all2g} (left panel) shows the fraction of second generation stars
remaining bound after gas expulsion, $f_\text{2G} = N_2/(N_1+N_2)$,  as a
function of the fraction of
remaining bound stars, $f_\text{bound}=(N_1+N_2)/N_\text{ini}$. All cases are located between two extreme scenarios: no
loss of second generation stars (upper dashed line) and no preferential
loss of first generation stars (horizontal dashed line). All the cases
computed by \citet{BaumgardtKroupa2007} that have more than 100 bound
stars at the end of the integration are presented.  To assess the
statistical significance of the results we estimate the statistical error
from $\sigma =
\sqrt{\sigma_1^2 + \sigma_2^2}$, where $\sigma_1=N_1^{-0.5}$ and $\sigma_2=
N_2^{-0.5}$ are
the statistical uncertainties for the number of first and second generation stars. In
clusters which do not lose many stars (points on the right), the error is
dominated by $\sigma_2$, while $\sigma_1$ and $\sigma_2$ contribute
significantly for clusters suffering a large loss of stars. For clusters
with less than 500 stars the remaining error is about 20\%.

A general trend is clearly visible in that the more dissolved clusters also
have a higher fraction of second generation stars at the end. This behaviour is
expected since second generation stars are initially more bound to the cluster
and the gas expulsion occurs on a timescale short compared to the
two-body relaxation timescale. For clusters near disruption, the fraction of
second generation stars can be as high as 70-75\% although with poor
statistics. N-body simulations with a higher number of initial stars will be
very helpful to quantify these numbers more precisely.

Even if the fraction of second generation stars is a monotonic function
  of the fraction of stars remaining bound to the cluster, other cluster
  properties 
  present a more pronounced dependence with the initial
  parameters. Figure~\ref{fig:all2g} (right panel) shows the final
  half-mass radius of the cluster as a function of the fraction of stars
  remaining bound to the cluster. For clusters with high $f_\text{bound}$
  the half-mass radius increases with decreasing value of the star formation
  efficiency. Indeed clusters with a high star formation efficiency are less
  perturbed when the primordial gas expulsion happens due to a lower mass
  of the gas remaining after star formation. 
  This trend roughly remains at low $\epsilon$ value but with
  more scatter.

\subsection{Initial conditions for globular cluster formation}

\begin{figure*}
\includegraphics[width=\textwidth]{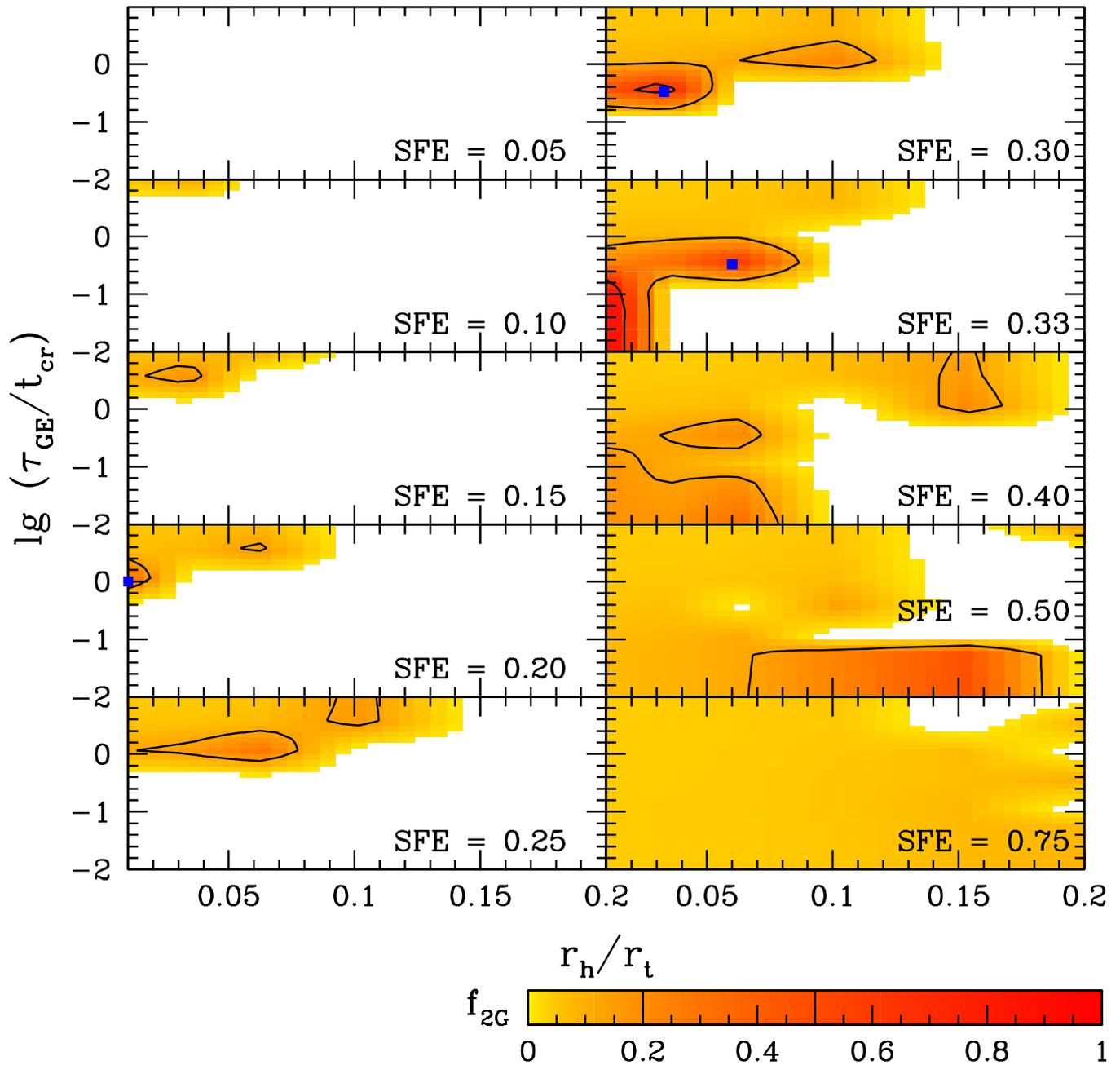}
\caption{Fraction of second generation stars, $f_\text{2G}$, which remain
  bound after gas expulsion as a function of various initial
  parameters. Each panel has a different value for the star formation
  efficiency (from 0.05 to 0.75).  White areas indicate dissolved clusters
  (i.e., with less than 100 bound stars). Black lines indicate levels where
  $f_\text{2G}$ is equal to 20\% (outer contour) and 50\% (inner
  contour). Blue dots in panels SFE=0.20, 0.30 and 0.33 indicate the three
  models presented in Fig.~\ref{fig:lossGE} (Note that the ratio of
    the gas expulsion timescale to the crossing time is expressed in
    decimal logarithm).}
\label{fig:alllev}
\end{figure*}

Figure \ref{fig:alllev} shows the fraction of second generation stars that
remain bound to the cluster after
the end of the gas expulsion phase for all the input parameters used by
\citet{BaumgardtKroupa2007}. White areas indicate fully disrupted clusters
while yellow to red colours indicate the fraction of second generation stars at
the end of the computation.  By varying the SFE, we retrieve the three main
behaviours for clusters. For clusters with low SFE ($\eps${} between 0.05
and 0.1), almost no cluster can survive.  In contrast, gas expulsion has
almost no effect for high SFE cases (see bottom right panel with SFE of
$\eps=0.75$). In this last case $f_\text{2G}$ remains always lower than
20\%.

Intermediate values of the SFE around 0.3-0.33 are more interesting as many
cases lead to a high fraction of second generation stars still bound to the
cluster. For a SFE of 0.33, a high fraction of second generation stars is
obtained for concentrated clusters with a short timescale for gas expulsion
($\tgcr\le$1). These candidates could be good progenitors of real GCs (see
second and third cases in Fig.~\ref{fig:lossGE} in \S~3.2). It should be
noted that a value of 0.33 for the SFE is close to the one found by
\citet{ParmentierFritze2009} from their study of the mass evolution of
clusters and is also consistent with direct observational surveys
\citep{LadaLada2003}. As we have seen, clusters with a small \rht{}
  ratio are more prone to present a high fraction of second generation
  stars. Thus this fraction should be higher with large distance to the
  Galactic centre (larger $r_\text{t}$). Alternatively, the tidal field may
  have been weaker because the Galaxy was not yet assembled. The
  observational trend observed by \citet{Carretta2006} who shows that
clusters with large orbital period and with high orbital inclinations
relative to the Galactic plane produce more extended O-Na and Mg-Al
anticorrelations \citep[see also][]{FraixBurnetDavoust2009} could be
  the imprint of the primordial gas expulsion process.

In addition to a SFE around 0.33, other regions of parameter space favour 
a large increase of the fraction of second generation stars,
namely a SFE around 0.25 combined with a fast timescale for gas expulsion
and a very concentrated cluster ($\rht \le0.05$). A last possibility is for
a higher SFE ($\eps = 0.5$) and an initially extended cluster ($\rht
\ge0.1$). However this case produces too extended clusters compared to
the observed ones so that only SFEs around or below 0.33 are allowed
  to increase the fraction of second generation stars.

\section{Towards a complete scenario for the evolution of globular clusters}

As we have seen in the previous section, primordial gas expulsion can be a
very efficient mechanism to increase the ratio between second and first
generation stars. The most favourable physical conditions for globular
cluster formation and early evolution are: (1) a star formation
efficiency around $\eps=0.33$, (2) a concentrated cluster relative to the
tidal radius and (3) a fast timescale for gas expulsion. In the following
we would like to consider how these constraints can be used to refine the
scenario of the evolution of globular clusters with pollution by fast
rotating massive stars.

\subsection{The wind of fast-rotating massive stars scenario}

As already outlined in the introduction, our scenario requires some basic
assumptions that are detailed in \citet{DecressinCharbnnel2007}. Let us
recall here the main points. We suppose that the first stellar generation
contains stars with initial masses between 0.1 and 120~M$_{\odot}$
following a standard IMF with a Salpeter-like slope for stars more massive than
0.8~\Ms{} and a log-normal distribution for lower mass stars
(\citealp{ParesceDeMarchi2000}), but that second generation stars consist
only of low-mass long-lived stars with initial masses between 0.1 and
0.8~M$_{\odot}$\footnote{The assumption about the mass range of
  second-generation stars is made only in order to minimise the constraints
  on the mass budget.}. We consider that the first generation polluters are
fast-rotating massive stars (i.e., with initial masses above
$\sim$~25~M$_{\odot}$) that enrich the ISM through their slow mechanical
winds loaded with H-burning products. We assume mass segregation,
primordial gas
expulsion and long-term 
evaporation of first generation low-mass stars in order to reproduce the
large fraction of second generation long-lived stars we see today (see \S~3).

The evolution of GCs passes through different key phases:
\begin{enumerate}
\item \textit{Formation of a first generation of stars}. First generation
  stars (over the complete mass range 0.1 to 120~M$_{\odot}$) form from a
  giant molecular cloud with a ``normal'' chemical composition similar to
  that of contemporary halo field stars of similar metallicity. As
  shown in \S~3, specific conditions are required at that phase: (1)
  an initial star formation efficiency for first generation stars defined
  as the total mass enclosed within first generation stars relative to the
  initial mass of the proto-cluster cloud in the same volume, $\eps[1G]$,
  around 0.3 -- 0.33 and (2) an initially highly concentrated cluster with a
  small half-mass radius (up to a few pc, see \S~4.2.4).

\item \textit{Evolution of fast-rotating massive stars ($m>25$~\Ms) and cluster
    pollution}.  Part of the pristine gas that has not been consumed to
  form first generation stars must sit within the cluster during the
  lifetime of the less massive polluters, i.e., $\sim$ 7-10~Myr for a
  25~\Ms{} star (see \S~4.2.4 below). Indeed the Li-free matter ejected by
  massive stars in their slow winds has to be mixed with pristine Li-rich
  intra-cluster gas in order to explain the Li-Na anticorrelation observed
  in NGC~6752 \citep{PasquiniBonifacio2005} and 47~Tuc
  \citep{BonifacioPasquini2007} and NGC~6397 \citep{LindPrimas2009}. This
  anticorrelation can actually be used to constrain the amount of pristine
  gas involved in this dilution process (see \S~4.2.1).

\item \textit{Formation of second generation stars}. Second generation
  low-mass stars (0.1 - 0.8~M$_{\odot}$) form from the ISM material
  polluted to various degrees by the slow winds of massive stars loaded
  with H-burning products. They have to be more centrally concentrated than
  their first generation counterparts, as required by the number ratios
  between first and second generation objects we observe today.
  \citet{DecressinCharbnnel2007} propose that massive polluters of the
  first generation could be born in the centre of the cluster or could have
  migrated there rapidly through mass-segregation. In both cases, the second
  generation stars are created in their immediate vicinity and share a
  similar radial distribution. Here we define the formation efficiency of
  the second stellar generation, $\eps[2G]$, as the ratio of the total mass
  enclosed by second generation stars to the total mass of the slow winds
  ejected by massive stars and the ISM matter used for the dilution
  process. 

\item \textit{Gas expulsion by SN}. Then gas expulsion occurs and removes
  the interstellar gas left after the two episodes of star
  formation\footnote{Note that in principle second generation massive stars
    could lead to the formation of a third generation of stars, but the higher
    order generations would comprise only an insignificant fraction of the
    whole population.}. For this process to be efficient enough the gas expulsion timescale
  must be very short, i.e., in the explosive regime, $\tau_\text{GE}<
  t_\text{cr}$ (see \S~3). We propose
  that this process is induced by the supernova explosions of the first
  generation stars that did not contribute to the chemical pollution, i.e.,
  with initial masses below 20--25~\Ms{} (the more massive progenitors
    implode see \S~4.2.3 below). During this phase most of the
  first generation stars that occupy the outer regions of the cluster are
  lost into the Galactic halo while second generation stars are more
  centrally concentrated and remain bound to the GC.

\item \textit{Long-term dynamical evolution}. Later, the long-term
  relaxation-driven evaporation leads to the preferential loss of
  first generation stars over 2-3 relaxation timescales (see
  \citetalias{DecressinBaumgardt2008}). Finally, both populations are
  mostly mixed and no further evolution of the number ratio between first
  and second generation long-lived stars is possible
  (\citetalias{DecressinBaumgardt2008}). Nowadays only the globular cluster
  $\omega$~Cen keeps a memory of the different initial
  distributions between first and second generations stars, because the
  two-body relaxation time in the core is comparable to the cluster age.

\end{enumerate}

\subsection{Possible issues}

The scenario outlined above faces several issues that require further discussion.

\subsubsection{Total star formation efficiency}

The first issue is related to the dilution process between ISM and the
ejecta of the polluters, and more specifically to the amount of pristine
gas consumed to form second generation stars. Indeed this process may
change the total SFE, which is the main parameter affecting the efficiency
of gas expulsion as discussed in \S~3.3.

\citet{DecressinCharbnnel2007} determined that in order to reproduce the Li-Na
anticorrelation in NGC~6752, the mass ratio between pristine gas and slow
stellar winds is around 1.15 after integration over time and IMF of the
massive star polluters.  On the other hand the mass lost by massive stars
and recycled into the second generation represents only 3.5-4\% of the
total mass of the first generation stars.  Thus for a star formation
efficiency for the first generation of $\eps[1G]=0.33$, and assuming that
all the matter ejected in the slow winds of the polluters is converted
into stars with a dilution factor of 1.2 with pristine gas,\footnote{These
  assumptions are needed to maximise the initial number of second
  generation stars.}  we find that at most 1.32\% of the protocluster gas
is used to form the second sellar generation.  Thus the SFE is only
slightly modified (by a few percent) by the second episode of star
formation.  In other words, the SFE is determined mainly by the formation of the
first stellar generation.

\subsubsection{Initial mass of proto-GC clouds}

We can now try to estimate the initial mass of the proto-gas clouds from
which globular clusters are born. From the global value of the SFE (around
0.33) within the cluster forming volume the proto-gas cloud is about three
times more massive than the total mass of created stars of first and second
generation. To explain the high fraction of anomalous stars observed in
present-day globular clusters (around 85\% in NGC~6752
\citealp{DecressinCharbnnel2007}, 70\% in NGC~2808
\citealp{PrantzosCharbonnel2006}) a large fraction of the stars born in the
cluster should have been lost from the cluster during its
evolution. \citet{DecressinCharbnnel2007} estimate that around 95\% of
first generation stars need to be lost in NGC~6752. Thus the initial mass
of NGC~6752 should be at least 10 times more massive than its present-day mass. Given
the luminosity of NGC~6752 \citep{Harris1996} and a mass-to-light ratio of
$M/L_\text{v} = 3$, we evaluate its actual mass to be $\sim3\times
10^5$~\Ms{}. Thus the mass of the proto-gas cloud should have been of the order of
$9\times 10^6$~\Ms{}.

However NGC~6752 is one of the most extreme cases in the number of anomalous
stars. \citet{CarrettaBragaglia2008} statistically study 19 GCs and find
that stars with abundance anomalies (their intermediate and extreme
populations) represent 50 to 80\% of cluster stars. Therefore NGC~6752 can
be one of the most massive GCs initially, and the
initial mass of most GCs is only of order of several $10^6$~\Ms{}. These high stellar
masses would explain why the pre-supernova feedback energy is not able to
expel the gas from the cluster \citep{BaumgardtKroupa2008}.

\subsubsection{Detailed chronology: early cluster evolution}

\begin{figure}
  \includegraphics[width=0.5\textwidth]{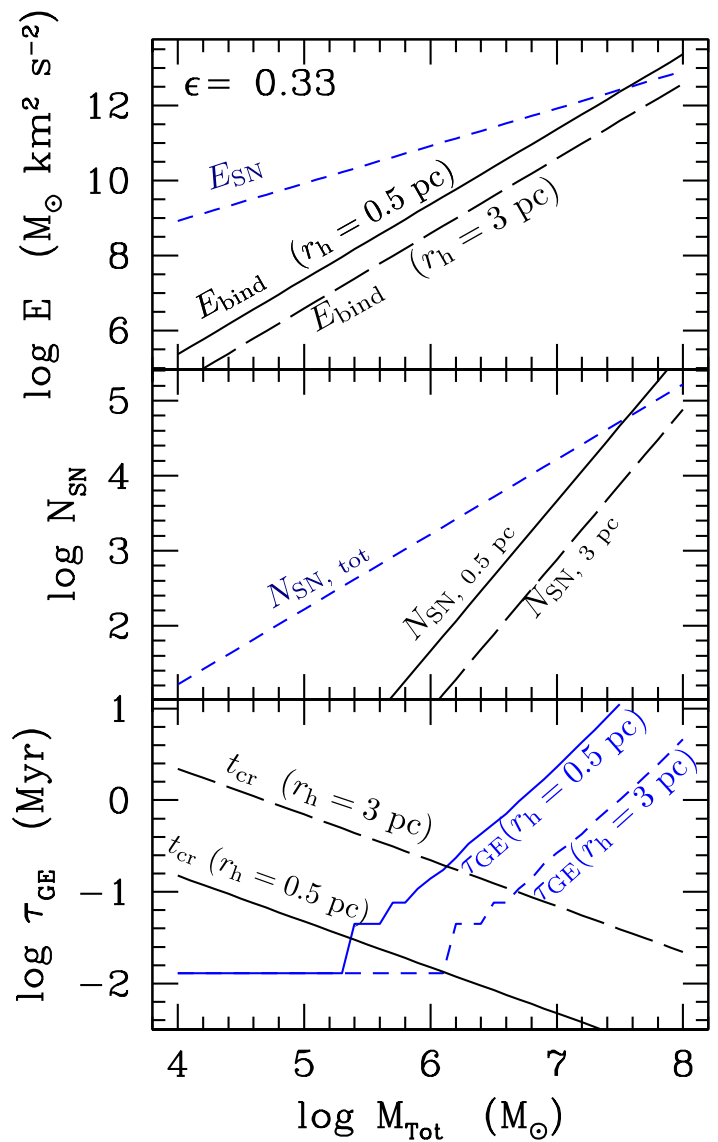}
  \caption{\textit{Top:} binding energy of a gas cloud as a function of its
  mass with an initial half-mass radius of 0.5~pc (full line) and 3~pc
  (long-dashed lines).The total
  energy released by SN for stellar progenitor masses between 8 and
    25~\Ms{}
is indicated as the short-dashed line. 
  \textit{Middle:} number of SNe needed to unbind a cluster as a
  function of its total mass with an initial half-mass radius of 0.5~pc (full
  line) and 3~pc and the total number of SNe created by the cluster (in the mass
  range 8-25~\Ms{}).
  \textit{Bottom:} Crossing-time of a cluster as a
  function of its total mass with an initial half-mass radius of 0.5~pc (full
  line) and 3~pc (long-dashed line) and corresponding gas expulsion timescale
  (full line and short-dashed line, e.g., \citealp{BaumgardtKroupa2008}).}
  \label{fig:snge}
\end{figure}

\begin{figure}
  \includegraphics[width=.5\textwidth]{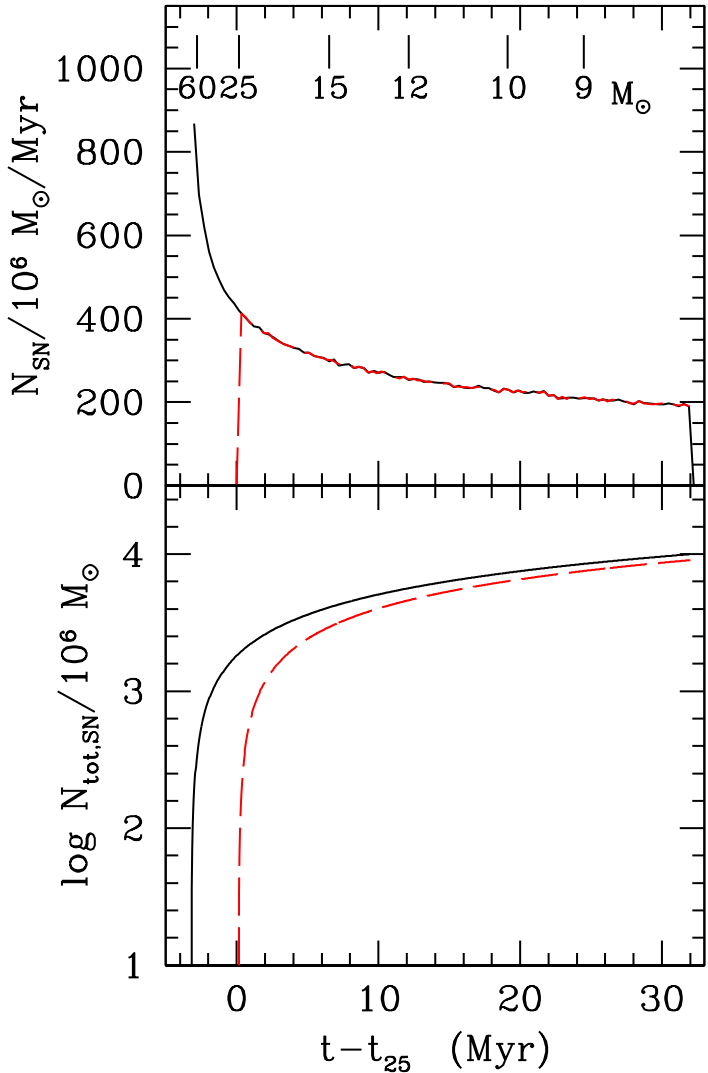}
  \caption{\textit{Top:} supernova rate as a function of time, using the
    \citet{HurleyPols2000} stellar-evolution routines, for a single stellar
    population of $10^6$~\Ms{} following a canonical IMF. Full
    and dotted lines refer to cases where all stars above 8~\Ms{} produce a
    SN, or only stars in the mass range 25-8~\Ms{},
    respectively. \textit{Bottom:} integration over time of the number of
    SNe. In both panels time is shifted to the SNe produced by stars
    with an initial mass of 25~\Ms{}.}
  \label{fig:SNR}
\end{figure}

One key issue concerns the necessity to retain pristine gas within the
cluster during the lifetime of the polluters in order to account for the
presence of lithium in the atmosphere of second generation stars. This
constraint is relevant for the issue whether very massive stars which
produce the first SN are unable to clear out the cluster from its gas.

This condition can also be fulfilled if at the end of their lives the
polluter stars do not undergo supernova explosions but rather directly
collapse into black holes, avoiding injection of SN kinetic energy into
the ISM.  Observational clues from black hole X-ray binaries
\citep{PortegiesZwartVerbunt1997,ErgmavandenHeuvel1998} as well as
nucleosynthesis constraints \citep{Maeder1992,KobulnickySkillman1997}
suggest that black holes form from stars with masses above $\sim$
25~M$_{\odot}$ \citep[see also][]{HegerFryer2003}. This rough limit is
confirmed by 2D core-collapse SN simulations \citep{Fryer1999},
which also predict that progenitor stars more massive than $\sim$ 40~M$_{\odot}$
are unable to launch shocks and thus do not produce a SN explosion.
In view of the sensitivity of core collapse simulations to the (uncertain)
input physics, these numbers must be considered with caution.  For example,
lowering the mean neutrino energy by 20 $\%$ lowers the fallback black hole
limit to $\sim$ 15~M$_{\odot}$ \citep{Fryer1999}. In addition, rotation
is expected to decrease the lower mass limit for black hole formation
\citep[e.g.,][]{HirschiMeynet2004}. Besides, 
\citet{GeorgyMeynet2009} found that the lowest masses which allow black hole
formation decrease from 40 to 30~\Ms{} when the metallicity decreases from
solar to $Z=0.004$. 

Despite the theoretical uncertainties regarding the formation of black holes and the physics of SN explosions, 
what matters at this level of the discussion is the fact 
that the lower mass limit for a star to collapse directly into a black hole
is very close to that of the fast-rotating massive polluter stars. 
This provides a natural way to avoid the deposition of 
SN kinetic energy into the ISM for the first 7-10~Myr of GC evolution, 
and thus to retain pristine gas within the cluster during the lifetime of the polluters.

\subsubsection{Detailed chronology: onset of gas expulsion}

On the other hand, if all massive stars with initial mass below 25~\Ms{} end
up as SN we still have to check if they are numerous enough to expel
the remaining pristine gas from the cluster potential
well. Fig.~\ref{fig:snge} (top panel) compares the binding energy for
clusters with initial half-mass radii of 0.5 and 3~pc assuming a SFE of
0.33 with the energy released by all SNe for stars between 25 and 8~\Ms{} as
a function of the cluster initial mass (gas and stars). These quantities
are computed following \citet{BaumgardtKroupa2008}. In the middle panel, we
show the number of SNe needed to unbind the cluster as well as the total
number of SNe produced from stars in the mass range 25--8~\Ms{}.
 
Both the binding and SN energies increase
with the cluster mass but with different rates: the binding energy scales as
$M_{Tot}^2$ while the SN energy scales linearly. For clusters with mass
below $10^{7.5}$~\Ms{} (i.e., the mass of proto-GC gas clouds), SN from
stars in the mass
range 8--25~\Ms{} produce
enough energy for gas expulsion to operate. 

\subsubsection{Fast gas expulsion}

Another issue concerns the timescale for gas expulsion, which has to be fast
compared to the crossing time according to our scenario.

To determine the timescale of gas expulsion by massive stars in the range
8--25~\Ms{} we present in Fig.~\ref{fig:SNR} the SN rates we expect from a
stellar population following a canonical IMF \citep{Kroupa2001}
normalised to a $10^6$~\Ms{} star cluster. We also indicate the total
number of SNe the cluster has over time (bottom panel). After 30~Myr
around 5000 SNe have exploded and even within 1~Myr after the 25~\Ms{} stars
explode about 1000 SNe are produced.

We can now compare the time at which enough SNe are produced to unbind a
cluster to the crossing time of the cluster.  This result is shown in
Fig.~\ref{fig:snge} (bottom panel) where both timescales are indicated. For
clusters with small half-mass radii (around 0.5~pc), only clusters with
initial mass below a few $10\time 10^5$~\Ms{} experience fast gas
expulsion as needed. However such a mass is too low to be the initial mass of
GCs. On the other hand, clusters with large half-mass radii (around 3~pc)
can produce fast gas expulsion up to a mass of $5\times10^6$~\Ms{}, which is
consistent with the value we infer for most proto-globular clusters. In
the extreme case of NGC~6752 (with mass up to $9\times10^6$~\Ms{}, see
\S~4.2.2) the initial half-mass radius should be of the order of 3--5~pc to
allow a fast enough gas expulsion. 

This confirms the results of \citet{ParmentierFritze2009}, who show that for
clusters with mass around $10^6-10^7$~\Ms{} gas expulsion is more likely
to happen in 2 or 3 crossing times implying an adiabatic regime (i.e., small
radius). Besides our findings are mainly compatible with the ones of
  \citet{MarksKroupa2008,MarksKroupa2010} which explain the relation
  between the slope of the
  actual mass function and the concentration of globular clusters as the
  dynamical response for the gas expulsion process. However the initial
  mass and radius of the cluster found by \citet{MarksKroupa2010} are
  smaller than the ones obtain in this paper as they deduced a top-heavy
  IMF in order to obtain gas expulsion.
 
\subsubsection{Place of birth of second generation stars}

Let us now address the issue of how the slow winds of fast rotating massive
stars are recycled into second generation
stars. \citet{DecressinMeynet2007} assume that the matter inside the
equatorial disc can already start to condense and produce a proto-stellar
object. However this local formation of second generation stars cannot
allow the formation of a distinct main sequence as observed in $\omega$~Cen
\citep{BedinPiotto2004} and NGC~2808 \citep{PiottoBedin2007} as pointed out
by \citet{DecressinCharbnnel2007} and \citet{Renzini2008}.

However it is possible that the strong radiation pressure accelerates the
disc so that it dissipates on larger scales. Here we assumed that the disc
is dense enough so that stellar radiation is not able to accelerate the
matter above the escape velocity of the cluster. The matter originating
from the disc will be stored outside the cluster centre and will fall back
when it is cold enough. The main difficulty of this scenario is to prevent
the mixing between SN ejecta with these slow winds. This can be the case if
the ISM (and the diluted slow winds) present inhomogeneities such that
SN ejecta escape the cluster mainly through low-density regions creating
tunnels \citep[see][]{PrantzosCharbonnel2006}. A similar view is
  presented by \citet{PalouvsWunsch2009}: thermal instabilities developed
  when the energy deposition to the ISM is dominated by stellar winds (with
  H-burning products). A large part of the matter sinks into the cluster
  centre in the form of compact high-density and cold gas which can be used
  to form second generation stars. Latter when SNe dominate the input
  energy, the cluster is cleared out by a stationary outflow \citep[see
  also][]{TenorioTagleWunsch2007,WunschTenorioTagle2008}.

In this case it will be possible that the slow component (\ie{} the slow
winds enriched with H-burning matter) remains decoupled from the SN
ejecta and cools down so that it migrates towards the cluster centre. There
the pressure of the ISM increases and second generation stars form. In
this case second generation stars will have a much more homogeneous
chemical composition compared to the stochastic formation process near
their massive progenitors. Therefore this scenario may reproduce the
discrete He-sequences inferred from observations. However the continuous
O-Na distribution found in many clusters is still challenging
\citep[see e.g.,][]{CarrettaBragaglia2009}. 

\section{Conclusions}

In this paper we have studied the influence of primordial gas expulsion by
supernovae during the early dynamical evolution of globular clusters.
in the
context of clusters with two chemically and dynamically distinct stellar
populations. 
In particular we investigate if this dynamical process can explain
  the high number of observed second generation stars which harbour abundance anomalies in light
  elements. We deduce the following:
\begin{itemize}
\item If the two populations have a different radial extent with second
  generation stars more concentrated, primordial gas expulsion is able to expel most of
  the first generation stars while most second generation stars can be 
  retained.
\item For a given fractional mass loss by the cluster, the fraction of second
  generation stars is nearly
  independent of the gas expulsion parameters (see \S~3.2).
\item The final observed fraction of second generation stars can constrain
  the initial properties of GCs as this fraction is highest for clusters with
  SFE around 0.3-0.33, with concentrated clusters relative to the tidal
  field, and with a fast timescale for gas expulsion relative
  to the crossing time.
\item We infer proto-GC cloud masses of several
  $10^6$~\Ms{} and up to $9\times 10^6$~\Ms{} for clusters which show a
  large fraction of chemically different second generation stars like
  NGC~6752. Their initial half-mass radii are in the range of $\sim1$--3~pc
  (4--5~pc for the most massive cases).
\item It is possible to reproduce 
  the fraction of second generation stars in present-day GCs through
  cluster dynamical processes by combining gas expulsion and tidal stripping
  during long-term evolution of initially mass-segregated clusters.
\item The primordial gas expulsion process can also be at the origin of
    the observational trend observed by \citet{Carretta2006}, who shows that
    clusters with large orbital period and with high orbital inclinations
    relative to the Galactic plane produce more extended O-Na and Mg-Al
    anticorrelations.
\end{itemize}

\begin{acknowledgements}
T.D. and C.C. acknowledge financial support from the french Programme National de
Physique Stellaire (PNPS) of CNRS/INSU, and from the Swiss National Science
Foundation (FNS). 
\end{acknowledgements}

\bibliographystyle{aa}
\bibliography{BibADS}

\end{document}